\documentclass[12pt]{article}

\usepackage{amsmath}
\usepackage{amsfonts}
\usepackage{amsthm}
\usepackage[usenames]{color}
\usepackage{setspace}
\usepackage{ulem,lipsum}
\usepackage{mathrsfs}
\usepackage{amssymb}

\usepackage{sgame}
\usepackage[citecolor=blue, colorlinks]{hyperref}
\usepackage[margin=3cm]{geometry}
\usepackage{natbib}
\usepackage{graphics}

\theoremstyle{definition}

\theoremstyle{definition}

\theoremstyle{plain}

\theoremstyle{definition}
\newtheorem{defn}{Definition}
\theoremstyle{definition}

\theoremstyle{plain}
\newtheorem{thm}{Theorem}
\theoremstyle{plain}
\newtheorem*{thm*}{Theorem}
\theoremstyle{plain}

\theoremstyle{plain}

\theoremstyle{definition}

\theoremstyle{remark}

\theoremstyle{remark}

\newcommand{\red}[1]{{\color{red}#1}}
\newcommand{\blue}[1]{{\color{blue}#1}}
\newcommand{\green}[1]{{\color{green}#1}}
\newcommand{\magenta}[1]{{\color{magenta}#1}}

\newcommand{\squareset}[1]{\left[#1\right]}

\newcommand{\set}[1]{\left\{#1\right\}}
\newcommand{\Set}[1]{\Bigl\{#1\Bigr\}}
\newcommand{\SET}[1]{\Bigg\{#1\Bigg\}}

\newcommand{\comment}[1]{}

\newcommand{\ceiling}[1]{\lceil#1\rceil}
\newcommand{\Ceiling}[1]{\Big\lceil#1\Big\rceil}

\newcommand{\Real}{\mathbb{R}}

\newcommand{\Natural}{\mathbb{N}}

\newcommand{\NA}{N^{A}}
\newcommand{\NB}{N^{B}}

\newcommand{\lovely}{m}
\newcommand{\modest}{\ell}

\newcommand{\UA}{U^{A}}
\newcommand{\UB}{U^{B}}

\newcommand{\nAlovely}{n_{A\lovely}}
\newcommand{\nBlovely}{n_{B\lovely}}

\newcommand{\nAl}{n_{A\lovely}}
\newcommand{\nAm}{n_{A\modest}}
\newcommand{\nBl}{n_{B\lovely}}
\newcommand{\nBm}{n_{B\modest}}

\newcommand{\nAs}{n_{As}}
\newcommand{\nBr}{n_{Br}}

\newcommand{\goodA}{\gamma_{A}}
\newcommand{\goodB}{\gamma_{B}}
\newcommand{\goodK}{\gamma_{K}}

\newcommand{\distaste}{\delta}
\newcommand{\distastestar}{\distaste^{\star}}

\newcommand{\rhostar}{\rho^{\star}}

\newcommand{\goodAstar}{\goodA^{\star}}

\newcommand{\goodArho}{\goodA^{\rhostar}}
\newcommand{\goodBrho}{\goodB^{\rhostar}}

\newcommand{\omegall}{\omega_{\lovely\lovely}}
\newcommand{\omegalm}{\omega_{\lovely\modest}}
\newcommand{\omegaml}{\omega_{\modest\lovely}}
\newcommand{\omegamm}{\omega_{\modest\modest}}

\newcommand{\omegat}{\omega^{t}}

\newcommand{\OmegaAprefl}{\Omega^{A, \lovely\succ \modest}}
\newcommand{\OmegaAprefm}{\Omega^{A, \modest\succ \lovely}}
\newcommand{\OmegaBprefl}{\Omega^{B, \lovely\succ \modest}}
\newcommand{\OmegaBprefm}{\Omega^{B, \modest\succ \lovely}}

\newcommand{\nAlAstar}{\nAl^{A^{*}}}
\newcommand{\nBmAstar}{\nBm^{A^{*}}}
\newcommand{\nAmBstar}{\nAm^{B^{*}}}
\newcommand{\nBlBstar}{\nBl^{B^{*}}}

\newcommand{\khat}{\hat{k}}

\newcommand{\Gcal}{\mathcal{G}}

\newcommand{\Ncal}{\mathcal{N}}

\newcommand{\Games}{\mathbb{G}}

\newcommand{\gammaA}{\gamma_{A}}
\newcommand{\gammaB}{\gamma_{B}}
\newcommand{\gammaK}{\gamma_{K}}

\newcommand{\GAA}{G^{AA}}
\newcommand{\GAB}{G^{AB}}
\newcommand{\GBB}{G^{BB}}

\newcommand{\nAsBstar}{\nAs^{B^{*}}}
\newcommand{\nBrBstar}{\nBr^{B^{*}}}

\newcommand{\sh}{s_{h}}
\newcommand{\si}{s_{i}}
\newcommand{\sj}{s_{j}}
\newcommand{\sk}{s_{k}}

\newcommand{\Ui}{U_{i}}

\newcommand{\omegastar}{\omega^{\star}}

\newcommand{\pibeta}{p_{i}^{\beta}}

\newcommand{\mubeta}{\mu^{\beta}}

\begin{document}

{\title{\Large 
Tiebout sorting in online communities}}


\author{John Lynham\protect\footnote{\emph{Email}: \href{mailto:lynham@hawaii.edu}{lynham@hawaii.edu}; \emph{Web}: \href{http://www2.hawaii.edu/~lynham/}{http://www2.hawaii.edu/$\sim$lynham/}. \emph{Address}: Department of Economics, Saunders Hall 532, University of Hawai`i at M\=anoa, 2424 Maile Way,
Honolulu, HI 96822} \hspace{.4in} Philip R. Neary\protect\footnote{\emph{Email}: \href{mailto:philip.neary@rhul.ac.uk}{philip.neary@rhul.ac.uk}; \emph{Web}: \href{https://sites.google.com/site/prneary/}{https://sites.google.com/site/prneary/}. \emph{Address}: Department of Economics, Royal Holloway, University of London, Egham, Surrey, TW20 0EX.}}


\date{\today}

\maketitle

\vspace{.3in}

\begin{abstract}
\noindent
This paper proposes a stylized, dynamic model to address the issue of sorting online.
There are two large homogeneous groups of individuals. 
Everyone must choose between two online platforms, one of which has superior amenities (akin to having superior local public goods).
Each individual enjoys interacting online with those from their own group but dislikes being on the same platform as those in the other group.
Unlike a Tiebout model of residential sorting, both platforms have unlimited capacity so there are no constraints on cross-platform migration.
It is clear how each group would like to sort themselves but, in the presence of the other type, only the two segregated outcomes are guaranteed to be equilibria.
Integration on a platform can be supported in equilibrium as long as the platform is sufficiently desirable.
If online integration of the two communities is a desired social outcome then the optimal policy is clear:\ make the preferred platform even more desirable.
Revitalizing the inferior platform will never lead to integration and even increases the likelihood the segregation.
Finally, integration is more elastic in response to an increase in platform amenities than to reductions in intolerance.
\end{abstract}



\newpage

\setstretch{1.3}

\section{Introduction}\label{INTRODUCTION}

\begin{quote}
	``Twenty-four hours of scrolling through posts from “Truthsayers” on the two-year-old platform explained why the site is tanking. In short, partisan echo chambers are stale, musty spaces that lack the sort of oppositional views needed to make social media tick. Truth Social feels like a MAGA town hall in a ventless conference room, where an endless line of folks step up to the mic to share how the world is out to get them.''\footnote{Lorraine Ali, ``I spent 24 hours on Trump’s Truth Social so you don’t have to. No wonder it’s tanking'', \textit{Los Angeles Times}, April 3, 2024.}
\end{quote}

Online segregation is a relatively new phenomenon. The rise of the internet, smart phones, and social media has created a world wherein everyone, but especially younger generations, are interacting more and more with their peers online. Survey evidence suggests that British teenagers now spend more time online than they do outdoors.\footnote{See, for example, \cite{frith2017social} and ``The average child spends just 7 hours a week outside, but more than twice that playing video games inside'', \emph{The Daily Mail}, 24 July, 2018.}
The hope that social networks and online communities would lead to more interconnectedness has, at least for now, been replaced by the fear that online platforms are actually more polarized and segregated than physical ones \citep{boyd2017america}.
It has been documented that internet users seek out spaces that they expect will be populated by people they identify with \citep{mcilwain2017racial}, which may partly explain why social media usage is often segregated not only by race and ethnicity, {but also by political inclination and overall worldview} \citep{duggan2015demographics}. For example, the ``white flight'' of the 1950s and '60s \citep{boustan2010postwar} was paralleled in the late 2000s exodus from ``ghetto'' Myspace to ``elite'' Facebook \citep{boyd2013white} and the more recent migration of academics from Twitter to Mastodon \citep{kupferschmidt2022musk,jeong2023exploring}.\footnote{We note that empirical work in the economics literature focused on online news consumption has found both ``no evidence that the Internet is becoming more segregated over time'' (\cite{gentzkow2011ideological}, p. 1799) and ``evidence for both sides of the debate'' (\cite{flaxman2016filter}, p. 298).}

In this paper we introduce a model designed to address the issue of online sorting.
We build a large population, simultaneous move game with two online platforms and two types of player.
There is consensus as to which platform is superior (e.g., sleeker interface, faster speeds, better chat features, video editing tools, guaranteed anonymity, etc.).
Each individual enjoys interacting online with those from their own group (``allies'') but dislikes being on the same platform as those in the other group (``trolls'').
We suppose that the large population interacts in this way indefinitely -- modelled using evolutionary dynamics -- and we show how seemingly innocuous micro-level individual decisions can lead to surprising macro-level population outcomes.
The key departure from the existing literature on sorting and segregation in physical neighborhoods is as follows:\ there are no capacity constraints nor congestion effects so everyone is always free to switch platform at any time if they so wish.\footnote{The assumption of no congestion effects for online platforms seems reasonable to us. For example, you probably did not notice that 500,000 new users joined Facebook today. This would not be true for the physical neighborhood that you live in.}

The one-shot version of our model always possesses at least two equilibria:\ the two segregated outcomes that are mirror images of one another.
The groups have differing preferences over these two Pareto efficient outcomes since every player prefers the outcome where their group resides on the optimal platform and the other group is out of sight and out of mind.
For certain parameters, integration on either platform can also be supported in equilibrium but this requires the benefit to coordinating with those from one's own group (on the superior platform) to trump the distaste experienced from interacting with those from the other group.
In essence, integration being supported in equilibrium places a demand on the quality of the hosting platform.

Having classified the equilibrium set, we then suppose that our model is the stage game of a repeated interaction.
We introduce population dynamics and show that every equilibrium is a rest point and only equilibria can be rest points.
To remove the equilibrium multiplicity, we employ the equilibrium selection technique of stochastic stability \citep{FosterYoung:1990:TPB,Young:1993:E}.
This resolves the two-dimensional race between ``location fundamentals'' and endogenous platform composition. 
We interpret the selected equilibrium as the most likely equilibrium to emerge given sufficient time, and we perform a variety of comparative statics in order to explore what nudges online communities towards being more or less integrated.

Some of our findings are surprising.
Perhaps obvious upon reflection, but, at least to us, not evident from the outset of this project.
For example, suppose that online integration is a desired social outcome.
Then the optimal policy in our setting is clear:\ make the desirable platform even more desirable.
In particular, improving the less desirable platform (the online equivalent of gentrification) never leads to integration.
In fact, if the world is already integrated then a policy based on platform revitalisation might lead to resegregation.
It is also possible to foster integration by reducing intolerance.
However, we show that, at least in terms of marginal elasticities, achieving integration through improving desirable platforms is more cost effective than attempting to promote integration by reducing intolerance.\footnote{A limitation of our model is that tolerance is not endogenously determined: integration could lead to greater tolerance \citep{billings2021long}.}

While agents in our model are choosing between online communities and not physical neighborhoods, at its core our model is about how large populations sort \`{a} la \cite{Tiebout:1956:JPE}.\footnote{Tiebout's work has inspired a vast literature on sorting in housing markets including \cite{benabou1993workings}, \cite{epple1991mobility}, \cite{epple1998equilibrium}, \cite{epple1999estimating}, \cite{de1990equilibrium}, and \cite{durlauf1996theory}. \cite{benabou1993workings}, for example, explores occupational segregation that occurs due to local complementarities in human capital.} 
Tiebout’s original motivation was the efficient delivery of local public goods, where sorting (i.e., segregating) like-minded households by type will occur in equilibrium and where such this occurrence is a good thing. 
Sorting in our model does not occur along fiscal dimensions as in Tiebout since there is no cost associated with locating on a platform.
Viewed in this way, our model is also related to the club theory approach of \cite{Buchanan:1965:Economica}, since the direct benefit associated with a specific good or service is dependent on the size of the consumption group.
However, the ``more the merrier'' property comes with the caveat that it holds only as long as the ``more'' are like-minded; every additional member of the other group who join the platform is not desirable as it reduces the online experience.
That is, not only is the size of the club important, but the make-up of the club is too.

Tiebout's findings are mathematically simple but stark.
Adapting these insights to our framework, Tiebout shows how groups of like-minded individuals will sort themselves because, in the event that certain people deem themselves better off elsewhere, then, using Tieboutian terminology, they will ``vote with their feet'', and relocate to the other site.
In fact, if our model had only one group, then Tieboutian dynamics would immediately predict where that group of like-minded individuals would end up:\ the platform that they all unanimously prefer.
The real novelty of our model is that we include a second group of individuals (who have the same preferences over platforms) coupled with no capacity constraints on either platform.
Since individuals also care about individual types at the same location, our model also speaks to the segregation models developed by \cite{Schelling:1969:AER, Schelling:1971:JMS, Schelling:1978:}.\footnote{Related papers include \cite{Akerlof:1997:E}, \cite{Arrow:1998:JEP}, \cite{BrockDurlauf:2001:RES}, \cite{Clark:1991:D}, \cite{IoannidesSeslen:2002:EL}, \cite{Krugman:1996:}, \cite{LindbeckNyberg:1999:QJE}, \cite{Manski:2000:JEP}, \cite{PancsVriend:2007:JPE}, \cite{Rosser:1999:JEP}, \cite{SkyrmsPemantle:2009:}, and \cite{Zhang:2004:JEBO,Zhang:2004:JMS,Zhang:2011:JRS}.}
With preferences over neighbors and not just neighborhoods, the allure of a superior site might no longer be enough to induce a group to sort where they would if left to their own devices.
In essence, our work injects distaste for others into a Tiebout style dynamic evolutionary framework to explore the question of sorting and segregation in online communities.

In contrast to many large populations models of sorting that hinge on tipping points, an interesting feature of our model is the existence of ``tipping sets''.
This generalizes the notion of a tipping point in that there is no single ratio of types that defines the individual threshold rule for switching platforms.
Rather, it is both the ratio of types on each platform coupled with the absolute numbers on each platform that determine when switching platforms is optimal.
One novel aspect of tipping sets is that the platforms can be tipped from integrated to segregated by a random shock that leaves the ratio of groups on a platform unchanged.
In fact, it is even possible that those who left would have preferred the ratio of types on the platform after the shock while those who remain preferred the ratio before.
These are two relatively unique predictions of our model that, to the best of our knowledge, remain untested empirically. 

The closest model to ours, in a purely mathematical sense, is the language game of \cite{Neary:2011:MGG,Neary:2012:GEB}.
In that model, everyone in a large population of two homogenous groups chooses from a common binary-action choice set.
The difference is that in the language game an individual values coordinating with every other individual, and does so to a measure that is independent of the other person's group affiliation.
Within-group interactions are symmetric coordination games with different preferred outcomes, and the across-group interaction is a battle of the sexes.
In the model of this paper, both within-group interactions are symmetric coordination games, although the across-group interaction is an anti-coordination game.
This is a richer set up since platform choice in our model is, adopting game-theoretic jargon, a strategic complement with one's own type and a strategic substitute with those from the other group.
The differences in the two models can easily be gleaned from comparing the three pairwise interactions on page 305 of \cite{Neary:2012:GEB} with those in Figure~\ref{fig:pairwiseGames} of this paper.\footnote{Both the language game and the model of this paper are particular examples of a multiple-group game as defined in \cite{Neary:2011:MGG}. A multiple group game is in fact a special case of a polymatrix game as defined in \cite{Janovskaya:1968:}.}

The paper closest to our own, in an economic sense, is \cite{BanzhafWalsh:2013:JUE} that considers the link between place-based investments and neighborhood tipping.\footnote{See also the related work of \cite{sethi2004inequality}.}
In their model, individuals choose physical locations and have preferences both over demographic composition and location-based public goods.\footnote{In fact, their model is rich enough to allow for heterogeneity within each type.}
The key difference is that the platforms in our model (the equivalent of neighborhoods) do not have capacity constraints.\footnote{Not restricting ourselves to settings with capacity constraints means that our model is a better fit for many online communities, such as virtual worlds, free dating sites, chat-boards, forums, and social media networks.}
While capacity constraints for locations seems intuitive for physical neighborhoods and maybe even some online platforms, it has the consequence that every assignment is a Nash equilibrium when the measure of agents equals the measure of slots - the reason being that every slot other than one's own is occupied so there is nowhere to move to.\footnote{Chapter 1 of \cite{Young:2001:} describes a game-theoretic model of segregation that circumvents this issue by allowing cooperative behaviour. Specifically, individuals are connected via a social network, and, despite the total number of individuals equalling the total number of slots, relocation is possible because pairs of players can ``swap'' homes when it is Pareto improving to do so.}
For this reason, the \cite{BanzhafWalsh:2013:JUE} model is a general equilibrium model and not game-theoretic.
Moreover, their model is static and always has at least two equilibria, and so there is an equilibrium selection problem.
By using dynamic equilibrium selection techniques, we remove this issue and so can study welfare properties of the selected equilibrium and perform comparative statics on it.

The balance of the paper is as follows.
Section~\ref{EXAMPLE} presents a particular instance of our model and solves it to completion.
We begin by emphasizing the important role played by platform quality, and we illustrate how straightforward the sorting problem is when the population is homogeneous.
We believe the big picture idea of our model can be understood by a quick skim of this section and we encourage a time-constrained reader to adopt this approach.\footnote{For example, we show how to classify the equilibrium set and the predicted equilibrium, and also provide some guidance on how to interpret and visualize tipping sets.}
Section~\ref{MODEL} defines the static model and solves for the equilibrium set.
In Section~\ref{DYNAMICS}, we introduce dynamics and use them to classify tipping sets.
The main result of Section~\ref{SELECTION} determines which equilibrium will be selected in the long run, while that of Section~\ref{COMPARATIVESTATICS} shows how the selected outcome varies with various changes in features of the two groups.
Section~\ref{CONCLUSION} concludes.

\section{An example of sorting online}\label{EXAMPLE}

There is a group of $17$ identical internet users that we refer to as Group $A$.
There are two available social media platforms, given by the set $\set{\ell,m}$, that each individual must choose from (we make the assumption that individuals are constrained in that they do not have time to exist on more than one platform).
Since the individuals are like-minded, they value being on the same platform as others.
This allows them to exchange ideas, swap stories, interact socially, and so on.
But, of course, this requires the individuals to coordinate on the same platform.

While coordination is good, let us assume that platform $m$ is superior to $\ell$ (the `$\ell$'  and `m' stand for `less' and `more' desirable respectively) so that any pair of Group $A$ individuals, labelled as $A_{1}$ and $A_{2}$, interact via the following pure coordination game.
\begin{center}
\begin{game}{2}{2}[$A_{1}$][$A_{2}$]\label{}
 	& $\ell$	& $m$\\
$\ell$	& $0.16, 0.16$ & $0, 0$ \\
$m$ 	& $0, 0$ & $0.84, 0.84$
\end{game}
\end{center}

\bigskip

\noindent
The game above has two pure strategy equilibria, $(\ell, \ell)$ and $(m, m)$.
As regards the population environment as a whole, we assume that each player's utility is the sum of their payoffs from interacting pairwise with everyone else.
Restricting attention to pure strategies means that the population coordination problem has two pure strategy equilibria in which all $17$ users adopt a common platform, either $\ell$ or $m$.
Clearly the equilibrium where everyone adopts platform $m$ is optimal.

The game described above is static.
Let us now suppose that this environment becomes the stage game of a recurrent interaction.
We assume that every so often individuals are afforded the opportunity to switch platform, and we model this using evolutionary dynamics.
Put yourself in the position of one of these individuals and ask yourself what platform you would choose.
With only two possible platforms, optimal behavior is described by a simple threshold rule:\ choose $m$ if at least {3} others are there and platform $\ell$ otherwise (i.e., when {15} or more others are on it). 
Given that the population is homogeneous, between 2 and 3 users on platform $m$ becomes the tipping point for society.\footnote{While we have ruled out mixed strategies, in part because the mixed strategy equilibria are extremely unstable in large population models, we note that the tipping point above is computed using the mixed strategy equilibrium of the two player game $\GAA$.}
We note that while behavior can (in theory) tip, in fact, under the current assumptions it never will, since whatever platform is currently optimal for one player is optimal for all.
Initial conditions matter enormously:\ if 3 or more players start out on platform $m$, then evolutionary pressure to the equilibrium in which all choose $m$ wins out; otherwise, everyone ends up on platform $\ell$.

With two possible ways in which the population might sort itself, there is an equilibrium selection problem.
The equilibrium multiplicity is resolved by the concept of stochastic stability \citep{FosterYoung:1990:TPB, Young:1993:E,KandoriMailath:1993:E}.
The basic idea is that individuals are not infallible optimizers, but rather occasionally make poor decisions. 
In this example, that would mean choosing to exist on the platform that is not currently optimal when afforded the opportunity to revise your decision.
The consequence of allowing for suboptimal behaviour is that equilibria can be escaped from because suboptimal choices can be followed by further ones.
Loosely put, the problem reduces to computing how likely it is that each of the equilibria will be escaped from and then evaluating the relative likelihood of these rare events. 
In the example of Group $A$ above, escaping the equilibrium where everyone chooses platform $\ell$ is far more likely since only three individuals relocating to platform $m$ is enough for everyone else to want to follow suit.
So, by employing stochastic stability, we expect the population to sort themselves optimally; in fact they will sort themselves optimally in  precisely in the manner as would be predicted by \cite{Tiebout:1956:JPE}.

Suppose now that there is another group of 13 internet users, that we will refer to as Group $B$.
Any pair of individuals from this group, let's refer to them as $B_1$ and $B_2$, interact via the following pure coordination game.
\begin{center}
\begin{game}{2}{2}[$B_{1}$][$B_{2}$]\label{}
 	& $\ell$	& $m$\\
$\ell$	& $0.05, 0.05$ & $0, 0$ \\
$m$ 	& $0, 0$ & $0.95, 0.95$
\end{game}
\end{center}

\bigskip

\noindent
We observe that Group $B$ individuals have even stronger relative preferences for platform $m$ than those in Group $A$.
This means that, by an almost identical analysis to that given concerning Group $A$, left to their own devices we would expect Group $B$ to sort themselves on to platform $m$.

If each group existed in isolation, we have mapped out how each would sort.
But a user on a platform can, and often will, interact with everyone who exists on the platform and not only with those that he or she would like to engage with.
Accordingly, if a Group $A$ individual and a Group $B$ individual choose the same platform, then they will interact, and so it remains for us to specify how.
We capture interactions of this form by the following anti-coordination game in which the row player, $A_{i}$, is a representative individual from Group $A$, and the column player, $B_{j}$, is from Group $B$.
\begin{center}
\begin{game}{2}{2}[$A_{i}$][$B_{j}$]\label{}
 	& $\ell$	& $m$\\
$\ell$	& $-0.45, -0.45$ & $0, 0$ \\
$m$ 	& $0, 0$ & $-0.45, -0.45$
\end{game}
\end{center}

\bigskip

\noindent
The anti-coordination game above captures the idea that individuals experience distaste from interacting with those from the other group since the two pure strategy equilibria are when the two individuals choose differently.
We emphasize the modelling choice taken that this distaste is the same for both groups and is platform-independent.\footnote{Note further that ``successfully avoiding'' each other brings a utility of zero.}

Each individual's problem is not as simple as before, since platform choice is now a strategic complement with those in one's own group while at the same time being a strategic substitute with those of the other group.
In order to compute optimal behavior, an individual uses the two-dimensional summary statistic that gives the number of users on platform $m$ (and hence the corresponding numbers on $\ell$).

What are the equilibria for this set up?
It turns out that there are three:\ the two ``segregated'' outcomes with the groups on different platforms, and an ``integrated'' outcome where everyone is located on the superior platform, $m$.
The segregated outcomes are always Pareto efficient equilibria in our model.
To see this, observe that each is the best possible outcome for those in some group.
The integrated equilibrium with everyone on the superior platform $m$ is also Pareto efficient in this example, although it is easy to check that in general it need not be.
The outcome with everyone located on site $\ell$ can also be supported in equilibrium, but not in this example because platform $m$ is deemed so much better.
Clearly this outcome can never be Pareto efficient since it is welfare dominated by the outcome with everyone on platform $m$.

While classifying the equilibria is straightforward, when studying dynamics we need to consider optimal behavior at every possible outcome.
The model with $N$ players appears to be unmanageable since the strategy space has $2^N$ elements, but because everyone in a given group is identical, all that is needed to classify optimal behaviour is a two-dimensional summary statistic that provides the number of those from each group on each platform.
This can be depicted by a two dimensional state space, as in Figure~\ref{fig:mainexamplepreferences}.
The integer on the $x$-axis lists the number of Group $A$ individuals on platform $m$ and the $y$-axis does likewise for Group $B$ members on platform $m$.
Since the group sizes are fixed, a two-dimensional summary statistic also provides the number of each type on platform $\ell$.

The colour coding of the states helps to understand what is going on.
All equilibria are depicted by large circles: the segregated equilibrium favoured by Group $A$ in \red{red}, that favoured by Group $B$ in \blue{blue}, and the integrated equilibrium in \magenta{magenta}.
States that are colour coded the same as an equilibrium represent those from where the dynamics are unambiguous as to where they will come to rest. 

Consider now the lines drawn in Figure~\ref{fig:mainexamplepreferences}.
These separate states where optimal behaviour of a group shifts. 
Consider the states coloured \blue{blue} just north west of the blue line and let us compare these with the non-blue states just south east of the blue line.
The former define states at which platform $\ell$ is optimal for those in Group $A$ while the latter represents those where platform $m$ is best for Group $B$.
On either side of this line, evolutionary forces are pulling Group $A$ towards a different outcome. 
We refer to this blue line as a {\it tipping set}.
Let us make an observation on tipping sets by comparing state $(2,5)$ with state $(4, 10)$.
From the perspective of a Group $A$ member, the ratio of the types on platform $m$ is the same for both states, but the optimal response is not the same for both.

We conclude this section with a discussion of equilibrium selection, recalling that there are only three viable candidates.
One conjecture might be the following.
The greater Group $A$ numbers, 17 vs 13, will render their most preferred outcome as the long run prediction.
However, this simple theory is at odds with an equally simple view that Group $B$ individuals have a stronger relative preference for being on site $m$.
Our upcoming Theorem~\ref{thm:selectionclassificationRestated} classifies what will be the stochastically stable equilibrium for this environment.
One can think of the result as supplying a machine that reads in the parameters of any such problem and, upon command, produces the long run prediction.



\section{The strategic setting}\label{MODEL}

Section~\ref{ssec:model} presents the model.
Section~\ref{ssec:equilibria} classifies the set of equilibria.
In Section~\ref{ssec:individual}, we consider the environment from the perspective of an individual from each group. 
This analysis will prove useful when we introduce dynamics in Section~\ref{DYNAMICS}; in particular it will allow us to formally define tipping sets.

\subsection{The model}\label{ssec:model}

An instance of the model, $\Gcal$, is defined as the tuple $\set{ \Ncal, \Pi, S, \Games}$, where $\Ncal := \set{1, \dots, N}$ is the {\it population} of players; $\Pi := \set{A, B}$ is a {\it partition} of $\Ncal$ into two nonempty homogeneous {\it groups}, Group $A$ and Group $B$, of sizes $\NA$ and $\NB$ respectively $(\NA, \NB \geq 2)$; $S := \set{\ell, m}$ are the two available platforms, referred to as ``less desirable" and ``more desirable" respectively\footnote{An online platform or website could be more desirable for fairly straightforward reasons such as it doesn't crash very often or it's easier to share/edit videos. A platform could also be more desirable for more abstract reasons such as being cooler or guaranteeing user anonymity.}; $\Games := \set{\GAA, \GAB, \GBB}$ is the collection of {\it local interactions}, where $\GAA$ is the pairwise exchange between a player from Group $A$ and a player from Group $A$, etc.  The three local interactions are given in Figure~\ref{fig:pairwiseGames} as follows,

\begin{figure}[htb!]
\begin{center}
\[
\hspace{.6in} \GAA \hspace{2.5in} \GBB
\]
\begin{game}{2}{2}[$A_{1}$][$A_{2}$]\label{game:GAA}
 	& $\ell$	& $m$\\
$\ell$	& $1$-$\gammaA$, $1$-$\gammaA$ & $0, 0$ \\
$m$ 	& $0, 0$ & $\gammaA$, $\gammaA$
\end{game}
\hspace{.3in}
\begin{game}{2}{2}[$A_{1}$][$A_{2}$]\label{game:GAA}
 	& $\ell$	& $m$\\
$\ell$	& $1$-$\gammaB$, $1$-$\gammaB$ & $0, 0$ \\
$m$ 	& $0, 0$ & $\gammaB$, $\gammaB$
\end{game}
\vspace{.1in}
\[
\hspace{.6in} \GAB
\]
\begin{game}{2}{2}[$A$][$B$]\label{game:GAB}
 	& $\ell$	& $m$\\
$\ell$	& $-\distaste$, $-\distaste$ & $0, 0$ \\
$m$ 	& $0, 0$ & $-\distaste$, $-\distaste$
\end{game}
\end{center}
\caption{Pairwise interactions $\GAA, \GAB$, and $\GBB$.}
\label{fig:pairwiseGames}
\end{figure}
where, for each group $K \in \Pi$, the parameter $\gammaK$ captures the {\it good} that a player from Group $K$ receives from coordinating with someone from the same group, while $\delta$ captures the {\it distaste} from locating at the same venue as someone from the other group.
We assume that both $\gammaA, \gammaB \in (1/2, 1)$, so that, while an individual wants to coordinate with those in his group, he would prefer to do so on platform $m$.
We constrain $\distaste$ to be positive, meaning that different types actively dislike locating on the same platform as those in the other group. 
We note that the distaste is location independent.
These payoff parameters are designed to capture the idea that individuals prefer to interact with people similar to themselves, and not with those of a different political, religious, or ethnic background.

We assume that each individual can only exist on one platform at a particular moment in time.
Since the two groups are homogeneous, what matters is the number of individuals from each group on a given platform and not precisely who those individuals are.
Letting $\nAlovely$ ($\nBlovely$) denote the number of Group $A$ ($B$) individuals choosing platform $\lovely$, the relevant strategic information is fully described by the two-dimensional summary statistic that we refer to as the \emph{state} of play $\omega = \big( [\omega]_{A}, [\omega]_{B} \big) = \big(\nAlovely, \nBlovely\big)$. The set of all states is given by $\Omega := \set{0, \dots, \NA} \times \set{0, \dots, \NB}$. The model is closed by specifying utility functions, $\UA$ and $\UB$, that a given type receives from choosing a particular platform when the state is $\omega$. These are given by
\begin{equation}\label{eq:Usomega}
\begin{aligned}
\UA(\lovely; \omega) &:= \Big( \nAlovely -1 \Big) \goodA -  \nBlovely \distaste\\
\UA(\modest; \omega) &:= \Big( \NA - \nAlovely - 1 \Big) (1 - \goodA) - (\NB - \nBlovely) \distaste\\
\UB(\lovely; \omega) &:=  - \nAlovely \distaste + \Big( \nBlovely - 1 \Big) \goodB\\
\UB(\modest; \omega) &:=  - \Big( \NA - \nAlovely \Big) \distaste + \Big( \NB - \nBlovely - 1 \Big) (1 - \goodB)
\end{aligned}
\end{equation}

Before moving on, let us discuss some obvious limitations of the set up.
First of all, we have assumed that both groups view platform $\lovely$ as the most desirable (albeit with potentially differing intensity).
While one can tell stories as to why this might be the case, clearly it need not always hold that such a ranking would be shared by one and all.
Second, while choice of platform is a strategic complement with those from the same group and a strategic substitute with those from the other group seems like a natural way to model segregation, the payoff normalisation we have chosen is arguably suspect.
Specifically, it is not clear why the payoff from failing to coordinate with those in the same group is precisely equal to the payoff from successfully anti-coordinating with those in the other group (both are normalised to zero).\footnote{For example, if an agent had a very strong dislike for those in the other group, it is possible that existing on a different platform to someone from the other group could actually yield positive utility.}
Lastly, that utilities are linear in the number of individuals from each group who locate at a venue could certainly be challenged.  One could argue that convexity or concavity (or a step function as in Schelling) is more appropriate.
The reason for such a constrained parameterisation is simply to make the employing of evolutionary dynamics easier to handle.

In defence of the limitations discussed above, each is easily remedied by modifying parameters of the game.
To see how, we note that model of this paper is a particular {\it multiple-group game} as proposed in \cite{Neary:2011:MGG}.
Looking at the payoff matrices $\GAA, \GBB$, and $\GAB$ in Figure~\ref{fig:pairwiseGames}, the only constraint of a multiple-group game is that the within-group pairwise interactions, in this case $\GAA$ and $\GBB$, are symmetric.
But as long as this constraint on payoffs is adhered to, the modeller is free to vary the parameters to fit the environment.

Lastly, we have assumed that each individual's choice of platform is payoff-relevant to everyone.
This need not be the case.
That is, while we have modelled payoff interdependencies using a complete network, there may be some situations for which less dense networks better capture the societal structure.
Migrating the set up from a complete network to an arbitrary network structure is easily incorporated - we leave this exciting extension to future work.\footnote{To do so one would proceed along similar lines to \cite{Neary:2011:MGG} and \cite{NearyNewton:2017:JMID}, both of which show how to extend the language game of \cite{Neary:2012:GEB} from a complete network to arbitrary network structures.}

\subsection{Equilibria}\label{ssec:equilibria}

We refer to population behavior at states $(0, 0), (0, \NB), (\NA, 0)$, and $(\NA, \NB)$, denoted by $\omegamm$, $\omegaml$, $\omegalm$, and $\omegall$ respectively, as {\it group-symmetric}. At states $\omegall$ and $\omegamm$, everyone is using the same platform so the groups are integrated, while at states $\omegaml$ and $\omegalm$ the groups are segregated. An instance of the model, $\Gcal$, is fully parameterised by the 5-element set of parameters $\set{\NA, \NB, \goodA, \goodB, \distaste}$, and from here on we define a given instance of $\Gcal$ by its associated parameter set. The following classifies when the various group symmetric states are equilibria.\footnote{It will be shown, once we introduce dynamics, that group-symmetric equilibria are the only serious candidates for long-run behavior.}

\begin{thm}\label{thm:classification}
Of the group-symmetric states, 
\begin{enumerate}
\item\label{lmequilibrium}
the segregated state $\omegalm$ is always a strict equilibrium.

\item\label{mlequilibrium}
the segregated state $\omegaml$ is always a strict equilibrium.

\item\label{llequilibrium}
the integrated state $\omegall$ is an equilibrium if and only if
\begin{equation}
\goodA \geq \frac{\NB}{\NA - 1}\distaste \hspace{.1in} \text{and} \hspace{.1in} \goodB \geq \frac{\NA}{\NB - 1}\distaste \label{eqn:llequilibrium}
\end{equation}

\item\label{mmequilibrium}
the integrated state $\omegamm$ is an equilibrium if and only if
\begin{equation}
1 - \goodA \geq \frac{\NB}{\NA - 1}\distaste \hspace{.1in} \text{and} \hspace{.1in} 1 -\goodB \geq \frac{\NA}{\NB - 1}\distaste \label{eqn:mmequilibrium}
\end{equation}
\end{enumerate}
\end{thm}

The proof of Theorem \ref{thm:classification} is straightforward and so is omitted.
Each part amounts to checking when a deviation from the group-symmetric state under consideration is profitable.

The result is intuitive.
Parts \ref{lmequilibrium} and \ref{mlequilibrium} state that both fully segregated outcomes are always strict equilibria.
This is almost automatic since a deviation from either of these profiles means that an individual goes from locating with everyone in her own group and away from those in the other group, to locating with those in the other group and away from everyone in her own group.
Clearly such a deviation can never be beneficial.
Parts \ref{llequilibrium} and \ref{mmequilibrium} are also straightforward.
The entire population locating on the same platform can be an equilibrium only if the total benefit that an individual receives from interacting with those in her own group on that platform exceeds the total distaste from interacting with those of the other group on the same platform (because the alternative -- moving to the empty platform -- guarantees a pay-off of zero).
Since platform $\lovely$ is viewed by everyone as preferable to platform $\modest$, it is immediate that state $\omegall$ is an equilibrium whenever state $\omegamm$ is.

Suppose the distaste parameter is large. Then both segregated outcomes are strict equilibria but the integrated outcomes are not. As individuals become more tolerant of each other, i.e., as $\delta$ decreases, the integrated outcomes can emerge as stable outcomes.
Consider, for example, the diverging trends of racial tolerance in-person and online \citep{stephens2014cost}.
Survey and interviewed-based measures of racial intolerance in the US document a decline from the 1980s to the late 2000s \citep{bobo2001racial,bobo2012real} and this is correlated with rising racial integration in many major cities \citep{farley1978chocolate,farley1993continued,glaeser2012end,ellen2012pathways,baum2016accounting}.
Conversely, racial distaste measured using anonymous behavior on online platforms appears to be more prevalent \citep{stephens2014cost} and this has been linked to relatively higher rates of sorting along racial lines in online communities \citep{boyd2013white}.

\subsection{Understanding Individual Behavior}\label{ssec:individual}

Since individuals in the same group are faced with the same strategic problem, it is useful to partition the set of states, $\Omega$, into a preference map for how representative individual from either group would behave at each state.
To this end we define the following,
\begin{align}
\OmegaAprefl &:= \Set{\omega \in \Omega \, \big| \, \UA(\lovely; \omega) > \UA(\modest; \omega)}\label{eqn:OmegaAprefl}\\
\OmegaBprefl &:= \Set{\omega \in \Omega \, \big| \, \UB(\lovely; \omega) > \UB(\modest; \omega)}\label{eqn:OmegaBprefl}\\
\OmegaAprefm &:= \Set{\omega \in \Omega \, \big| \, \UA(\modest; \omega) > \UA(\lovely; \omega)} \label{eqn:OmegaAprefm}\\
\OmegaBprefm &:= \Set{\omega \in \Omega \, \big| \, \UB(\modest; \omega) > \UB(\lovely; \omega)}\label{eqn:OmegaBprefm}
\end{align}
In words, $\OmegaAprefl$ and $\OmegaBprefl$ are the sets of states such that, given the current platform choices of everybody else, a Group $A$ individual and a Group $B$ individual respectively {strictly} prefer existing on platform $\lovely$ to platform $\modest$.
Similarly but opposite for $\OmegaAprefm$ and $\OmegaBprefm$.

Consider the following parameters as an example: $\Gcal = (\NA, \NB, \gammaA, \gammaB, \delta) = (17, 13, 0.84, 0.95, 0.45)$.\footnote{This is the same example as that sketched informally in Section~\ref{EXAMPLE}.}
It can be checked that neither inequality in part \ref{mmequilibrium} of Theorem \ref{thm:classification} is satisfied and so the equilibrium set is $\set{\omegalm, \omegaml, \omegall}$. Since states $\omegalm$ and $\omegaml$ are always strict equilibria, we use them as reference points. We begin by considering how many individuals of the same type would have to switch platforms for any individual's best-response to change.

To get us started, let us suppose that population behavior is given by state $\omegaml$ and, from here, let us compute how many Group $A$ individuals are needed to switch to $\lovely$ in order for platform $\lovely$ to become the best-response for everyone in Group $A$.  Similarly, we consider how many Group $B$ individuals would have to relocate to platform $\modest$ for platform $\lovely$ to become the best-response for everyone in Group $A$. This can be formalised as follows. Letting $\Natural$ denote the natural numbers and $\Real$ denote the real line, for any $x \in \Real$, let $\ceiling{x} := \min\set{n \in \Natural \, | \, x \leq n}$. Now define the following,
\begin{align}
\nAlAstar &:= \min\SET{\NA, \Ceiling{(1 - \gammaA)\NA + (2\gammaA - 1) + \NB\delta}},\label{eqn:nAlAstar}\\
\nBmAstar &:= \min\SET{\NB, \Ceiling{\frac{(\NA - 1)(1- \gammaA)}{2\delta} + \frac{\NB}{2}}},\\
\nBlBstar &:= \min\SET{\NB, \Ceiling{(1 - \gammaB)\NB + (2\gammaB - 1) + \NA\delta}},\\
\nAmBstar &:= \min\SET{\NA, \Ceiling{\frac{(\NB - 1)(1- \gammaB)}{2\delta} + \frac{\NA}{2}}}.\label{eqn:nAmBstar}
\end{align}

In words, the integer $\nAlAstar$ is the minimum number of Group $A$ individuals that have to choose platform $\lovely$, when all Group $B$ players choose $\lovely$, in order for $\lovely$ to be the best-response for those from Group $A$. If the $\min\set{\cdot, \cdot}$ operator ``kicks in", it means that all members of a given group switching their platform is not enough for the best-response to change. We will see below that when the $\min\set{\cdot, \cdot}$ operator does kick in, it means that neither integrated outcome, $\omegall$ or $\omegamm$, is an equilibrium.

The above is made clearer with a picture of the state space.\footnote{The procedure follows along similar lines to that of \cite{Neary:2012:GEB}.}
Since the state space is two-dimensional, it can be depicted as an $(\NA+1) \times  (\NB + 1)$ grid with, for any state $\omega$, $\squareset{\omega}_{A} = \nAl$ on the horizontal axis and $\squareset{\omega}_{B} = \nBl$ on the vertical axis. Figure \ref{fig:mainexamplepreferences} shows the state space $\Omega$ as an $18 \times 14$ lattice, with $\squareset{\omega}_{A} \in \set{0, \dots, 17}$ on the horizontal-axis, and $\squareset{\omega}_{B} \in \set{0, \dots, 13}$ on the vertical-axis.  Each state is depicted by a circle. The state space $\Omega$ in Figure \ref{fig:mainexamplepreferences} is partitioned as $\set{\OmegaAprefm, \OmegaBprefm, \OmegaAprefl \cap \OmegaBprefl}$, with states colour coded according to the element of the partition in which they lie.  The set of \blue{blue} circles is $\OmegaAprefm$, while the \red{red} circles denote $\OmegaBprefm$.  At states given by hollow circles, the set $\OmegaAprefl \cap \OmegaBprefl$, both groups prefer platform $\lovely$.  These sets are defined by $(\nAlAstar, \nBmAstar, \nBlBstar, \nAmBstar) = (10, 10, 10, 10)$, calculated using equations \eqref{eqn:nAlAstar}-\eqref{eqn:nAmBstar}.  That is, $(\nAlAstar, 0) = (10, 13)$, $(0, \NB - \nBmAstar) = (0, 3)$, $(\NA - \nAsBstar, 0) = (7, 0)$ and $(\NA, \nBrBstar) = (17, 10)$. Equilibrium states are depicted by large circles, with \magenta{magenta} (the combination of red and blue) representing the fully integrated state. For now, ignore the green \green{X} and ignore the fact that certain hollow states are shaded magenta.

\begin{figure}[h]
\begin{center}
\begin{picture}(360,290)(0, 0)
\put(20,275){\blue{\circle*{10}}}
\multiput(20,95)(0,20){10}{\blue{\circle*{4}}}
\multiput(40,115)(0,20){9}{\blue{\circle*{4}}}
\multiput(60,135)(0,20){8}{\blue{\circle*{4}}}
\multiput(80,155)(0,20){7}{\blue{\circle*{4}}}
\multiput(100,175)(0,20){6}{\blue{\circle*{4}}}
\multiput(120,195)(0,20){5}{\blue{\circle*{4}}}
\multiput(140,215)(0,20){4}{\blue{\circle*{4}}}
\multiput(160,235)(0,20){3}{\blue{\circle*{4}}}
\multiput(180,255)(0,20){2}{\blue{\circle*{4}}}
\put(200,275){\blue{\circle*{4}}}

\put(20,85){\blue{\line(1,1){190}}}

\put(210,275){\magenta{\line(0,-1){70}}}
\put(210,205){\magenta{\line(1,0){150}}}

\put(170,15){\red{\line(1,1){190}}}

\multiput(20,15)(0,20){4}{{\circle{4}}}
\multiput(40,15)(0,20){5}{{\circle{4}}}
\multiput(60,15)(0,20){6}{{\circle{4}}}
\multiput(80,15)(0,20){7}{{\circle{4}}}
\multiput(100,15)(0,20){8}{{\circle{4}}}
\multiput(120,15)(0,20){9}{{\circle{4}}}
\multiput(140,15)(0,20){10}{{\circle{4}}}
\multiput(160,15)(0,20){11}{{\circle{4}}}
\multiput(180,35)(0,20){11}{{\circle{4}}}
\multiput(200,55)(0,20){11}{{\circle{4}}}

\multiput(220,75)(0,20){7}{{\circle{4}}}
\multiput(220,215)(0,20){4}{\magenta{\circle{4}}}
\multiput(240,95)(0,20){6}{{\circle{4}}}
\multiput(240,215)(0,20){4}{\magenta{\circle{4}}}
\multiput(260,115)(0,20){5}{{\circle{4}}}
\multiput(260,215)(0,20){4}{\magenta{\circle{4}}}
\multiput(280,135)(0,20){4}{{\circle{4}}}
\multiput(280,215)(0,20){4}{\magenta{\circle{4}}}
\multiput(300,155)(0,20){3}{{\circle{4}}}
\multiput(300,215)(0,20){4}{\magenta{\circle{4}}}
\multiput(320,175)(0,20){2}{{\circle{4}}}
\multiput(320,215)(0,20){4}{\magenta{\circle{4}}}
\multiput(340,195)(0,20){1}{{\circle{4}}}
\multiput(340,215)(0,20){4}{\magenta{\circle{4}}}
\multiput(360,215)(0,20){4}{\magenta{\circle{4}}}
\put(360,275){\magenta{\circle*{10}}}


\multiput(180,15)(0,20){1}{\red{\circle*{4}{150}}}
\multiput(200,15)(0,20){2}{\red{\circle*{4}{150}}}
\multiput(220,15)(0,20){3}{\red{\circle*{4}{150}}}
\multiput(240,15)(0,20){4}{\red{\circle*{4}{150}}}
\multiput(260,15)(0,20){5}{\red{\circle*{4}{150}}}
\multiput(280,15)(0,20){6}{\red{\circle*{4}{150}}}
\multiput(300,15)(0,20){7}{\red{\circle*{4}{150}}}
\multiput(320,15)(0,20){8}{\red{\circle*{4}{150}}}
\multiput(340,15)(0,20){9}{\red{\circle*{4}{150}}}
\multiput(360,15)(0,20){10}{\red{\circle*{4}{150}}}
\put(360,15){\red{\circle*{10}}}

\put(0, 12){0}
\put(0, 32){1}
\put(0, 52){2}
\put(0, 72){3}
\put(0, 92){4}
\put(0, 112){5}
\put(0, 132){6}
\put(0, 152){7}
\put(0, 172){8}
\put(0, 192){9}
\put(-3, 212){10}
\put(-3, 232){11}
\put(-3, 252){12}
\put(-3, 272){13}


\put(115, 111){\green{\textbf{X}}}

\put(17, -2){0}
\put(37, -2){1}
\put(57, -2){2}
\put(77, -2){3}
\put(97, -2){4}
\put(117, -2){5}
\put(137, -2){6}
\put(157, -2){7}
\put(177, -2){8}
\put(197, -2){9}
\put(215, -2){10}
\put(235, -2){11}
\put(255, -2){12}
\put(275, -2){13}
\put(295, -2){14}
\put(315, -2){15}
\put(335, -2){16}
\put(355, -2){17}

\end{picture}
\smallskip
\hspace{.2in} \caption{$\OmegaAprefm = \blue{\bullet}$. \hspace{.2in} $\OmegaAprefl \cap \OmegaBprefl = {\circ}$. \hspace{.2in} $\OmegaBprefm = \red{\bullet}$. Vertical axis indicates the number of B agents on the more desirable platform. Horizontal axis indicates number of A agents on the more desirable platform. Equilibrium states are depicted by large circles, with blue representing all of the B agents and none of the A agents on the more desirable platform, red representing all of the A agents and none of the B agents on the more desirable platform, and magenta representing full integration on the more desirable platform.}
\label{fig:mainexamplepreferences}
\end{center}
\end{figure}

We have included lines separating the elements of the preference partition.
As we will see formally in the next section, these lines give an intuition for how to define {\it tipping sets}. For example, consider the blue line and the state $(10, 13)$. At this state, everybody in the population prefers platform $\lovely$. However, suppose that one Group $A$ individual using $\lovely$ switched to platform $\modest$ such that population behavior is now given by state $(9, 13)$. At this new state, platform $\modest$ is preferable for everyone in Group $A$.
When we introduce best-response based dynamics in the next section, we will see that these two states lead to different outcomes:\ complete integration and complete segregation.

We now highlight a novel feature of out set up.
Consider state $(2, 5)$ in Figure \ref{fig:mainexamplepreferences}. At this state, all individuals in the population prefer platform $\lovely$ to $\modest$.
Now suppose that the numbers of both Groups on platform $\lovely$ are doubled. That is, consider the state $(4, 10)$. The ratio of types on platform $\lovely$ remains as before, and yet state $(4, 10) \in \OmegaAprefm$ so that $\modest$ is now the preferred platform of Group $A$ members.
We note that the ratio of the two types using platform $m$ has not changed and yet optimal behavior for those in Group $A$ has changed.
Note further that this feature can occur even if the ratio on platform $m$ strictly improves for Group $A$; this can be seen by comparing state $(2,5)$ with the state $(3, 7)$ or the state $(4, 11)$. At both these states, the Schelling ratio of platform $m$ is better for those in Group $A$, and yet platform $\lovely$ is a less attractive option.

This is quite a different result to Schelling and similar work.
It demonstrates that in a world where ``locations'' do not have capacity constraints, integrated communities can be tipped to segregated, {\`{a} la} Schelling, but without any change in the ratio of one group to the other or even when the ratio becomes more favourable towards the group that ends up leaving.
In other words, we observe a levels effect in preferences in addition to a ratio effect.
Thus, the view that an online platform has become ``too A" or ``too B" can be triggered by an increase in the absolute numbers of one group or the other, even though relative diversity has remained the same.

\section{Evolutionary dynamics and tipping sets}\label{DYNAMICS}

We now suppose that the above model is the stage game of a repeated interaction. Time is discrete, begins at $t = 0$, and continues forever. Each period, one individual is randomly selected (each with equal probability of $1/N$) to update their online platform choice. This one-at-a-time revision protocol is known as {\it asynchronous learning} \citep{BinmoreSamuelson:1997:JET, Blume:2003:GEB}. When afforded a revision opportunity, we assume that the chosen individual takes a best-response to the current state. This revision protocol and behavioral rule pair satisfies the standard evolutionary assumption that better-performing actions are no worse represented next period.

Since our interest is in long run outcomes, we wish to track population behavior as it evolves.
To assist in this we will exploit the fact that our model is a {\it potential game} \citep{ShapleyMonderer:1996:GEB}.
A game is a {potential game} if the change in each player's utility from a unilateral deviation can be derived from a common function, referred to as the game's {\it potential function}.

To see how our model is a potential game, we note that each of the pairwise interactions $\GAA, \GAB$, and $\GBB$ are potential games and as such, by a result in \cite{Neary:2011:MGG}, so is the model as a whole, with potential equal to the sum of potentials of each two-player interaction.
More specifically, $\Gcal$, is a potential game with potential function $\rhostar : \Omega \to \Real$, where for any state $\omega \in \Omega$,
\begin{equation}\label{eq:potential}
\rhostar(\omega) := \sum_{\begin{subarray}{c}i, j \in A\\ i \neq j \end{subarray}} \, \rho^{AA}(\si, \sj) + \sum_{i \in A, \, k \in B} \, \rho^{AB}(\si, \sk) +\sum_{\begin{subarray}{c}h, k \in B\\ h \neq k \end{subarray}} \, \rho^{BB}(\sh, \sk)
\end{equation}
with $\rho^{AA}$, $\rho^{AB}$, and $\rho^{BB}$ all real valued functions defined as in \eqref{eq:potentials}.
%
\begin{align}
\rho^{AA}(\modest, \modest) &= 1 - \gammaA & \rho^{AA}(\modest, \lovely) = 0 \nonumber\\
\rho^{AA}(\lovely, \modest) &= 0 & \rho^{AA}(\lovely, \lovely) = \gammaA \nonumber\\
\nonumber\\
\rho^{AB}(\modest, \modest) &= \delta & \rho^{AB}(\modest, \lovely) = 0 \nonumber\\
\rho^{AB}(\lovely, \modest) &= 0 & \rho^{AB}(\lovely, \lovely) = \delta \label{eq:potentials}\\
\nonumber\\
\rho^{BB}(\modest, \modest) &= 1- \gammaB &\rho^{BB}(\modest, \lovely) = 0 \nonumber\\
\rho^{BB}(\lovely, \modest) &= 0 & \rho^{BB}(\lovely, \lovely) = \gammaB \nonumber
\end{align}

Given that our model is a potential game, we can state the following.

\begin{thm}\label{thm:equilibriumlockin}
When the revision protocol is asynchronous learning and individuals behave according to myopic best-response, with probability 1 population behavior will come to rest at one of the pure strategy equilibria.
\end{thm}

\noindent
The proof of the above is almost immediate.
Since our model is a finite potential game, we can appeal to Lemma 2.3 in \cite{ShapleyMonderer:1996:GEB}.
Asynchronous learning with a best-response based updating rule will generate a so-called ``finite improvement path'' that must terminate at some pure strategy equilibrium.

That the only candidates for long run behavior in our model are the group-symmetric states can be understood intuitively by referring to Figure \ref{fig:mainexamplepreferences}.  From any of the states shaded blue, $\OmegaAprefm$, population behavior must drift towards equilibrium $\omegaml = (0, 13)$.  Similarly but opposite for any of the red states, $\OmegaBprefm$, from where population behavior will ultimately come to rest at $\omegalm = (17, 0)$.  So it remains to consider the evolution from the hollow states. From these, the dynamics are pushing ``up and to the right'', and depending on the order that players are randomly called upon to act, population behavior will ultimately drift to either a blue or red state or to equilibrium $\omegall$.

While Theorem \ref{thm:equilibriumlockin} states that an equilibrium will be reached no matter what state the population begins at, the initial state need not uniquely determine the final rest point.
For example, consider state $(5, 5)$ labelled by the green \green{X} in Figure \ref{fig:mainexamplepreferences}. From here, there are best-response based paths that lead to $\OmegaAprefm$ (blue), others that lead to $\OmegaBprefm$ (red), and yet more that lead to equilibrium $\omegall$ (magenta). Thus, there is not always ``path dependence'' associated with this dynamic.\footnote{Had we instead considered a best-reply dynamic where all dissatisfied individuals switch platform every period, then each state would be uniquely identified with a particular equilibrium outcome.}

In the next section we will consider equilibrium selection.
Before that we show how to define tipping sets.
Letting $\omegat$ denote the population state at time $t \in \Natural$. We begin by defining the basin of attraction \citep{Ellison:2000:RES} of each equilibrium as follows:

\begin{defn}
For a group symmetric equilibrium $\omegastar \in \Omega$, the {\it basin of attraction} of $\omegastar$, $D(\omegastar)$, is the set of initial states from which population behavior will converge to $\omegastar$ with probability one. That is,
\[
D(\omegastar) = \Set{ \omega \in \Omega \,  \Big| \, \text{Prob}\big[\set{\exists \, t' > 0 \text{ such that } \omegat = \omegastar, \, \forall t > t' \, \big| \, \omega^{0} = \omega }\big] = 1 }.
\]
\end{defn}

\noindent
{
As an example in Figure \ref{fig:mainexamplepreferences}, $D(\omegalm)$ is the set of \blue{blue} states, $D(\omegaml)$ is the set of \red{red} states, and $D(\omegamm)$ is the set of \magenta{magenta} states that are north, east, or north east of the state $(10, 10)$, including the state $(10, 10)$ itself.
}

The idea of a tipping set then follows naturally. A tipping set is the set of outermost states of a particular basin of attraction. That is, the set of states that are closest to (i.e., a distance of 1 from) those not contained in the basin of attraction, where our notion of close is the taxicab metric ($L^{1}$ distance) on $\Omega$. Formally, for any pair of states $\omega', \omega''$ we define taxicab distance $\ell: \Omega \times \Omega \to \Natural_{0}$ as $\ell(\omega', \omega'') = \big|[\omega']_{A} - [\omega'']_{A}\big| + \big|[\omega']_{B} - [\omega'']_{B}\big|$. We now define a tipping set for an equilibrium.

\begin{defn}\label{defn:Tipping}
For group symmetric equilibrium $\omegastar \in \Omega$, the {\it tipping set} of $\omegastar$, $T(\omegastar)$, is defined by
\[
T(\omegastar) = \Set{ \omega \in D(\omegastar) \,  \Big| \, \exists \, \omega' \not\in D(\omegastar) \text{ with } \ell(\omega, \omega') = 1}.
\]
\end{defn}

Going back again to Figure \ref{fig:mainexamplepreferences}, it is now straightforward to see the tipping sets. We have that $T(\omegaml)$ is the set of blue states that are just north west of the blue line, $T(\omegalm)$ is the set of red states just south east of the red line, and $T(\omegall)$ is the set of states that are directly north or directly east of the state $(10, 10)$ including the state $(10, 10)$ itself. We emphasize again that the ratio of types at each state in a tipping set need not be the same. In fact, as in our example, the ratio can be different at each state in a given tipping set.

\section{Equilibrium Selection}\label{SELECTION}

To select between the equilibria, we assume that individuals occasionally choose suboptimal platforms.
Such ``mistakes'' are standard in the literature on evolutionary game theory.
Specifically, with $\omegat$ denoting the population state at time $t \in \Natural$, if player $i$ is afforded an opportunity to revise his choice he chooses platform $\lovely \in S$ according to the probability distribution $\pibeta(\, m \, | \, \omegat \,)$, where for any $\beta > 0$,
\begin{equation}\label{eq:response}
\pibeta(\, \lovely \, | \, \omegat \, ) \hspace{.1in} := \hspace{.1in} \frac{\text{exp}\big( \beta \, \Ui(\lovely; \omegat) \big)}{\text{exp}\big( \beta \, \Ui(\lovely; \omegat) \big) + \text{exp}\big( \beta \, \Ui(\modest; \omegat) \big)}
\end{equation}
and chooses platform $\modest$ with probability $1-\pibeta(\, \lovely \, | \, \omegat \, )$.

The behavioral rule above is known as logit response.
It makes the assumption that the more painful a mistake, the less likely it is to be observed.\footnote{It is well-known since \cite{BerginLipman:1996:E} that exactly how mistakes occur can influence the set of stochastically stable equilibria. We have assumed logit mistakes, but other leading examples include uniform errors \citep{KandoriMailath:1993:E,Young:1993:E}, directed mistakes \citep{NaiduHwang:2010:EL,HwangNaidu:2024:JEEA}, and condition dependent errors \citep{BilanciniBoncinelli:2020:ET}. \cite{LimNeary:2016:GEB} and \cite{MasNax:2016:JET} are experimental studies designed to understand how individuals in large population behaviour behave suboptimally.}
As $\beta \downarrow 0$, this approaches uniform randomization over both actions.
When $\beta \uparrow \infty$, the rule approaches best-response. Asynchronous learning coupled with this perturbed best-response describes a stochastic process with a unique invariant probability measure $\mubeta$ on the state space $\Omega$. A result from \cite{Blume:1993:GEB} shows that as $\beta \uparrow \infty$, all the probability mass under $\mubeta$ accumulates on the states that maximize the potential function. That is, the states that maximize the potential are precisely the {\it stochastically stable} equilibria \citep{Young:1993:E}.

To compute the potential at each group-symmetric state, we view the game as one on a fully connected graph with players representing vertices and local interactions represented by edges. The number of edges on any fully connected undirected graph with $N$ vertices is $N \choose 2$. Similarly, the number of edges on any fully connected bipartite graph, as is the case with the subgraph connecting all pairs of players from groups $A$ and $B$, of size $\NA$ and $\NB$ respectively, is $\NA \times \NB$. The potential at each group-symmetric state is then given by
\begin{equation}\label{eq:comparison}
\begin{array}{lcccccc}
\rhostar\big( \omegall \big) & = & {\NA \choose 2} \cdot \goodA & + & 0 & + & {\NB \choose 2} \cdot \goodB\\
\rhostar\big( \omegalm \big) & = & {\NA \choose 2} \cdot \goodA & + & (\NA \times \NB) \cdot \distaste & + & {\NB \choose 2} \cdot (1 - \goodB)\\
\rhostar\big( \omegaml \big) & = & {\NA \choose 2} \cdot (1 - \goodA) & + & (\NA \times \NB) \cdot \distaste & + & {\NB \choose 2} \cdot \goodB\\
\rhostar\big( \omegamm \big) & = & {\NA \choose 2} \cdot (1 - \goodA) & + & 0 & + & {\NB \choose 2} \cdot (1 - \goodB)\\
\end{array}
\end{equation}
where the first term on the right hand side of each equality is the potential due to interactions among Group $A$ individuals, the second due to interactions across the groups, and the third due to interactions among Group $B$ individuals.

Theorem \ref{thm:selectionclassificationRestated} below classifies for what range of the benefit pay-off parameters, $\gammaA$ and $\gammaB$, each equilibrium is stochastically stable. We simplify the notation by defining the following:
\begin{align*}
\goodArho(\NA, \NB, \distaste) := \frac{\NB}{\NA - 1}\distaste + \frac{1}{2} \hspace{.4in} &\text{ and } \hspace{.4in} \goodBrho(\NA, \NB, \distaste) := \frac{\NA}{\NB - 1}\distaste + \frac{1}{2}, \\
f(\NA, \goodA) := {\NA \choose 2} \cdot (2\goodA - 1) \hspace{.4in} &\text{ and } \hspace{.4in} f(\NB, \goodB) := {\NB \choose 2} \cdot (2\goodB - 1).\\\
\end{align*}

\begin{thm}\label{thm:selectionclassificationRestated}
	The following classifies when each equilibrium is stochastically stable.
	\begin{enumerate}
		\item\label{mmNotStable}
		Integration on the inferior platform, $\omegamm$, is never stochastically stable.
		
		\item\label{llStableRestated}
		Integration on the superior platform, $\omegall$ is stochastically stable if and only if
		\begin{equation}\label{eq:(l,l)stableRestate}
		\goodA \geq \goodArho \hspace{.4in} \text{ and } \hspace{.4in} \goodB \geq \goodBrho
		\end{equation}
		
		\item\label{lmStableRestated}
		Segregated equilibrium $\omegalm$ is stochastically stable if and only if
		\begin{equation}
		f(\NA, \goodA) \geq f(\NB, \goodB) \hspace{.2in} \text{ and } \hspace{.2in} \goodB \leq \goodBrho
		\end{equation}
		
		\item\label{mlStableRestated}
		Segregated equilibrium $\omegaml$ is stochastically stable if and only if
		\begin{equation}\label{eq:mlStableRestated}
		f(\NA, \goodA) \leq f(\NB, \goodB) \hspace{.2in} \text{ and } \hspace{.2in} \goodA \leq \goodArho 
		\end{equation}
		
	\end{enumerate}
\end{thm}

\begin{proof}
	\begin{enumerate}
		\item
		By comparing the first and fourth equations in \eqref{eq:comparison} and observing that both $\goodA > \frac{1}{2}$ and $\goodB > \frac{1}{2}$, it is immediate that the potential at $\omegall$ is always strictly greater than at $\omegamm$. As such $\omegamm$ is never stochastically stable.
		
		\item
		Given that $\omegamm$ is never stochastically stable, parts \ref{llStableRestated}, \ref{lmStableRestated}, and \ref{mlStableRestated} all follow from comparison of the first three equations in \eqref{eq:comparison}. We prove only part \ref{llStableRestated} as the rest follow in a similar manner.\\
		For $\omegall$ to be stochastically stable we need that both
		\[
		\rhostar(\omegall) \geq \rhostar(\omegalm) \hspace{.4in} \text{ and } \hspace{.4in} \rhostar(\omegall) \geq \rhostar(\omegaml).
		\]
		Considering the first inequality, from taking the first and second expressions in \eqref{eq:comparison}, we get that
		\[
		{\NA \choose 2} \cdot \goodA  +  {\NB \choose 2} \cdot \goodB \geq {\NA \choose 2} \cdot \goodA + (\NA \times \NB) \cdot \distaste + {\NB \choose 2} \cdot (1 - \goodB),
		\]
		which after some rearranging yields the first inequality in \eqref{eq:(l,l)stableRestate}. Considering the second inequality, and taking the first and third expressions in \eqref{eq:comparison}, we get that
		\[
		{\NA \choose 2} \cdot \goodA  +  {\NB \choose 2} \cdot \goodB \geq {\NA \choose 2} \cdot (1 - \goodA) + (\NA \times \NB) \cdot \distaste + {\NB \choose 2} \cdot \goodB,
		\]
		which after some rearranging yields the second inequality in \eqref{eq:(l,l)stableRestate}.
	\end{enumerate}
\end{proof}

As can be seen from the two expressions in \eqref{eq:(l,l)stableRestate} above, the integrated outcome where all individuals located on platform $\lovely$ is stochastically stable if and only if both benefit payoff parameters, $\goodA$ and $\goodB$, are sufficiently large.
This is really a requirement on platform quality which can be interpreted as being about the quality of the club good as in \cite{Buchanan:1965:Economica}.
A slight rearranging of the first expression in \eqref{eq:(l,l)stableRestate} above yields that one necessary condition for $\omegall$ to be stochastically stable is that $(\NA - 1) (2\goodA - 1) \geq 2\NB \distaste$.  Note that the term $2\goodA - 1$ is simply the difference in benefit that each of two Group $A$ individuals earn from successful coordination with each other on platform $\lovely$ over platform $\modest$ \big(since $2\goodA - 1 = \goodA - (1-\goodA)$\big).

That the integrated outcome with everyone on the inferior platform, state $\omegamm$, is never stochastically stable is intuitive. At both integrated outcomes, the distaste experienced by the presence of those from the other group is equal, but the benefit to coordinating with those in your own group is greater on platform $\lovely$. This confirms that improving the less desirable platform never leads to integration on that platform. The only way integration could occur is if the less desirable platform was improved so much that it became the more desirable platform but then segregation could re-emerge with the two groups switching platform. This is the idea of gentrification simply leading to re-segregation. This result stands in contrast to work on physical neighborhoods by \cite{fernandez1996income} and others who find that neighborhood revitalisation can lead to welfare-enhancing integration.

For the purposes of this short paper, we do not delve into the details here, but it is straightforward to show the stochastically stable equilibria, as classified by Theorem \ref{thm:selectionclassificationRestated}, are both Pareto efficient and maximise utilitarian social welfare.

\section{Comparative Statics}\label{COMPARATIVESTATICS}

If one accepts stochastic stability as an accurate predictor of long run behavior, some natural questions spring to mind. Suppose a planner wanted to induce an integrated outcome, and suppose further that the planner can affect preferences, but that after altering preferences, the planner can do no more. That is, the planner can affect preferences but then must release the stochastic system, letting the dynamics take over.
For example, if the planner's goal is to guide society towards integration, would she be better off by increasing the benefit parameters, $\goodA$ or $\goodB$, or by reducing the distaste parameter $\distaste$? This section explores questions of this kind. Again, we take the position that the stochastically stable equilibrium will emerge and we consider how payoffs (from the perspective of the agents) vary with the equilibrium that is selected. 

\subsection{Varying Group Size}

We note that the threshold benefit payoff for Group $A$, $\goodArho$, is decreasing in $\NA$ and increasing in $\NB$ (a similar but opposite statement holds for $\goodBrho$). In the extreme, this means that the integrated outcome, $\omegall$, will not be stochastically stable if one of the groups is considerably larger than the other. At the margin, the issue is more complicated. If the entire population is located on an integrated platform, then increasing the size of Group $A$ ($B$) will weakly move the situation towards the segregated outcome that is preferred by Group $A$ ($B$). If population behavior is described by the segregated outcome least preferred by Group $A$ ($B$), then the effect of increasing $\NA$ ($\NB$) is ambiguous (though still beneficial for those in Group $A$). As described above, continually increasing the size of Group $A$ will mean that ultimately their preferred segregated equilibrium will be selected. The issue is whether the integrated outcome is ``passed through'' along the way. For some parameters the transition from one segregated outcome to the other is instant, while for other parameters the integrated outcome is visited. Precisely when each of these occurs is described in the following theorem.\footnote{We state the theorem only for an increase in the size of Group $A$. Similar but opposite statements occur for increasing the size of Group $B$.}

\begin{thm}\label{thm:changingGroupSize}
Consider the model $\Gcal := (\NA, \NB, \goodA, \goodB, \distaste)$ and a sequence of models $\set{\Gcal_{k}}_{k=0}^{\infty}$, where along the sequence, the size of Group $A$ is incrementally increased.
That is, define $\Gcal_0 = \Gcal$, and for each $k = 1, 2, \dots$, define $\Gcal_{k} := (\NA + k, \NB, \goodA, \goodB, \distaste)$.
\begin{enumerate}
\item\label{omegalmincreasingNA}
Suppose the segregated equilibrium $\omegalm$ is stochastically stable for model $\Gcal$. Then, $\omegalm$ is uniquely stochastically stable for all models in $\set{\Gcal_{k}}_{k=0}^{\infty}$.

\item\label{omegallincreasingNA}
Suppose the integrated equilibrium $\omegall$ is stochastically stable for model $\Gcal$. Then there exists an integer $\khat \in \Natural$ such that $\omegall$ remains stochastically stable for all models in $\set{\Gcal_{k}}_{k=0}^{\khat-1}$ and such that $\omegalm$ is stochastically stable for all models in $\set{\Gcal_{k}}_{k=\khat}^{\infty}$.

\item\label{omegamlincreasingNA}
Suppose $\omegaml$ is stochastically stable for model $\Gcal$ so that $f(\NA, \goodA) \leq f(\NB, \goodB)$ and $\goodA \leq \goodArho$. If, by incrementally increasing $\NA$ we get $\goodA > \goodArho$ while $f(\NA, \goodA) \leq f(\NB, \goodB)$ then $\omegall$ becomes stochastically stable and we reduce to case \ref{omegallincreasingNA}. If, by incrementally increasing $\NA$ we get $f(\NA, \goodA) > f(\NB, \goodB)$ while $\goodA \leq \goodArho$ then $\omegalm$ becomes stochastically stable and we reduce to case \ref{omegalmincreasingNA}.
\end{enumerate}
\end{thm}

\begin{proof}
	
	\begin{enumerate}
		\item
		It is clear from the expressions in \eqref{eq:comparison} that $\rhostar(\omegalm)$ increases more than $\rhostar(\omegall)$ and $\rhostar(\omegaml)$ for any incremental increase in $\NA$.
		
		\item
		Given $\omegall$ is stochastically stable, we know that $\goodA \geq \goodArho$ and $\goodB \geq \goodBrho$. Furthermore, we have that $\goodArho$ is decreasing in $\NA$, so the first inequality will always hold. Eventually, for some $\khat$, we have $\goodB > \goodBrho(\NA + \khat, \NB, \distaste)$. Given that $\goodA$ is still large enough, it cannot be that $\omegaml$ is stochastically stable by the second inequality in \eqref{eq:mlStableRestated}, and so $\omegalm$ must be stochastically stable. From here we are reduced to case \ref{omegalmincreasingNA}, so $\omegalm$ remains stochastically stable.
		
		\item
		Suppose $\omegaml$ is stochastically stable, meaning both inequalities in \eqref{eq:mlStableRestated} hold. Eventually, one of these inequalities will cease to bind. If the first inequality breaks, then we are reduced to case \ref{omegallincreasingNA}. If the second inequality breaks, then we are reduced to case \ref{omegalmincreasingNA}.
\end{enumerate}
\end{proof}


Theorem \ref{thm:changingGroupSize} raises some interesting issues.
From the perspective of the individual, an increase in the size of one's group is always weakly beneficial since either (i) the stochastically stable equilibrium will remain the same with an extra individual in your group, or (ii) the selected equilibrium will change to one that yields a higher utility.
From the perspective of the planner, things are not so clear.
Suppose a planner favors integration and yet the stochastically stable equilibrium is a segregated outcome. One thing the planner could do is to increase the size of the group on platform $\modest$ in the hope that their increased numbers will induce them to locate at $\lovely$ such that integration will occur. However, part \ref{omegamlincreasingNA} of Theorem \ref{thm:changingGroupSize} says that this may not always be possible. Specifically, it may be that by increasing the size of the group at $\modest$ that eventually an immediate flip occurs whereby the other segregated outcome becomes stochastically stable. This result is akin to what is known as ``white flight" in the US \citep{boustan2010postwar}. White residents leave cities as immigration leads to more integration, fleeing to the suburbs and rural areas. Segregation occurs again, except the segregated outcome is a reversal of how it was initially.
In fact, Theorem \ref{thm:changingGroupSize} shows that unlimited ``immigration'' of one group always eventually leads to online segregation.

\subsection{Varying Payoffs}

Again let us suppose that one of the segregated outcomes is stochastically stable but the planner would like to see integration occur. Suppose further that the planner has the ability to affect preferences in the following sense. She can increase the benefit that two agents from the same group would earn on platform $\lovely$. For example, 
adding additional features or making the platform more user-friendly. The other alternative for the planner would be to reduce the common distaste that individuals feel towards those in the other group.

Specifically, suppose the segregated outcome $\omegaml$ is stochastically stable and suppose the planner favours integration. Further suppose that the planner has two options:\ (i) increase the attractiveness of platform $\lovely$ for Group $A$ members by a fixed percentage, or (ii) decrease distaste $\distaste$ (i.e., increase tolerance) for all individuals by reducing it by the same percentage. As Theorem \ref{thm:changingPreferences} below shows, if integration is the goal, it is always better (in terms of the elasticity of responses) to make the desirable platform even more desirable than to reduce distaste.

\begin{thm}\label{thm:changingPreferences}
Suppose a segregated outcome is stochastically stable but the social planner favors integration. Suppose further that the social planner has the ability to change preferences either by reducing $\distaste$ by $x\%$ or by increasing by $x\%$ the benefit that two agents from the group currently located on platform $\modest$ get from coordinating at $\lovely$. Whenever reducing $\distaste$ by $x\%$ renders the integrated outcome $\omegall$ stochastically stable, increasing $\goodK$ by $x\%$ will also. But the reverse need not hold.
\end{thm}

\begin{proof}
	Without loss of generality we suppose that $\omegaml$ is stochastically stable so that the first inequality in \eqref{eq:(l,l)stableRestate} does not hold. In the hope of making this inequality bind, the planner can make one of the following two changes:
	\begin{align*}
		\goodA & \mapsto \goodAstar = (1+x)\goodA\\
		\distaste & \mapsto \distastestar = (1-x)\distaste
	\end{align*}
	It is immediate that increasing $\goodA$ to $\goodAstar = (1+x)\goodA$ is an $x\%$ increase in the range of values for $\goodA$. We now show that reducing $\distaste$ to $\distastestar = (1-x)\distaste$ does not induce as large an expansion of values for $\goodA$. A reduction in $x\%$ of $\distaste$ to $\distastestar$ changes the threshold value $\goodArho(\NA, \NB, \distastestar)$.
	We have that 
	\begin{align*}
		\frac{\goodArho(\NA, \NB, \distastestar)}{\goodArho(\NA, \NB, \distaste)} &= \frac{\frac{\NB}{\NA - 1} \distastestar + \frac{1}{2}}{\frac{\NB}{\NA - 1} \distaste + \frac{1}{2}}\\
		&= \frac{\frac{\NB}{\NA - 1} (1-x)\distaste + \frac{1}{2}}{\frac{\NB}{\NA - 1} \distaste + \frac{1}{2}}
	\end{align*}
	which is a reduction in distaste that is strictly less than $x\%$.
\end{proof}

Theorem \ref{thm:changingPreferences} demonstrates that integration is more elastic with respect to a 1\% change in $\goodA$ compared to a 1\% change in $\distaste$. 
{We are, of course, ignoring costs here.}
Large cost differences could obviously lead to the conclusion that increasing tolerance is a more cost effective way to achieve integration.

\section{Conclusion}\label{CONCLUSION}

We have shown in this paper that when individuals care about both the features and the group composition of their online communities, their preferences are reflected in patterns of segregation that are sometimes surprising.
In our model, a platform can tip from integrated to segregated even when the ratio of the two types using that platform remains unchanged. If integration is the desired social outcome then the optimal policy in our model is clear:\ make desirable online platforms or social networks even more desirable. Revitalizing less desirable platforms (think Myspace, Bebo, or EconSpark) never leads to integration; in fact, it can lead to resegregation.
Further, integration is more elastic in response to improving desirable platforms than attempting to reduce intolerance. Which option is ultimately more effective will depend on relative costs.

Our analysis is abstract enough to permit alternative interpretations. For example, 
we have alluded to the fact that our model could be used to study physical neighborhoods, especially those where congestion or capacity constraints are not binding. In terms of future work, the most obvious unanswered question is whether empirical evidence exists for our most interesting segregation result:\  the ratio of $A$ to $B$ remains the same but an increase in the absolute number of $B$s causes the $A$s to move to 
a different online platform.
In other words, in a world of online Tiebout sorting without capacity constraints, do numbers matter more than Schelling's canonical ratio?

\newpage 
\bibliographystyle{aer}
\bibliography{libraryARS.bib}


\newpage

\end{document}